\documentclass[12pt,letter]{article}
\pdfoutput=1
\usepackage{graphicx}
\usepackage{amsmath}
\usepackage{amssymb}
\usepackage{cancel}
\usepackage[left=1in,right=1in,top=1in,bottom=1in]{geometry}
\usepackage{setspace}
\usepackage{float}
\usepackage[super,sort&compress,comma]{natbib}
\usepackage[font={footnotesize}]{caption} 
\usepackage[labelfont=bf]{caption}
\usepackage{bm}
\usepackage[T1]{fontenc}
\usepackage{microtype}
\usepackage{subfigure}
\usepackage{chemformula}
\usepackage[version=4]{mhchem}
\usepackage{textcomp}
\usepackage{color}
\usepackage{mathtools}
\usepackage{gensymb}
\usepackage{esvect}
\usepackage{booktabs}
\usepackage{multirow}
\usepackage{indentfirst}
\usepackage{mathrsfs}
\usepackage{float}
\usepackage{hyperref}
\usepackage{xcolor}
\usepackage{booktabs}
\usepackage{amssymb} 
\usepackage{bm} 
\begin{document}

\newcommand{\mub}{\boldsymbol{\mu}}

\title{\textbf{Deep Learning of Proteins with Local and Global Regions of Disorder}}
\author{Oufan Zhang$^{1}$, Zi Hao Liu$^{2, 3}$, Julie D Forman-Kay$^{2,3}$, Teresa Head-Gordon$^{1,4}$}
\date{}
\maketitle
\noindent
\begin{center}
$^1$Kenneth S. Pitzer Theory Center and Department of Chemistry, University of California, Berkeley, CA, USA\\
$^2$Molecular Medicine Program, Hospital for Sick Children, Toronto, Ontario M5S 1A8, Canada\\
$^3$Department of Biochemistry, University of Toronto, Toronto, Ontario M5G 1X8, Canada\\
$^4$Departments of Bioengineering and Chemical and Biomolecular Engineering, University of California, Berkeley, CA, USA\\

corresponding author: thg@berkeley.edu
\end{center}

\begin{abstract}
\noindent
Although machine learning has transformed protein structure prediction of folded protein ground states with remarkable accuracy, intrinsically disordered proteins and regions (IDPs/IDRs) are defined by diverse and dynamical structural ensembles that are predicted with low confidence by algorithms such as AlphaFold and RoseTTAFold. We present a new machine learning method, IDPForge (\textbf{I}ntrinsically \textbf{D}isordered \textbf{P}rotein, \textbf{FO}lded and disordered \textbf{R}egion \textbf{GE}nerator), that exploits a transformer protein language diffusion model to create all-atom IDP ensembles and IDR disordered ensembles that maintains the folded domains. IDPForge does not require sequence-specific training, back transformations from coarse-grained representations, nor ensemble reweighting, as in general the created IDP/IDR conformational ensembles show good agreement with solution experimental data, and options for biasing with experimental restraints are provided if desired. We envision that IDPForge with these diverse capabilities will facilitate integrative and structural studies for proteins that contain intrinsic disorder, and is available as an open source resource for general use.

\end{abstract}

\newpage
\section{Introduction}
\label{sec:intro}
\noindent
Structures of biomolecules have driven functional insight ever since Watson and Crick solved the structure of DNA, establishing the structure-function paradigm that has been critical to progress in molecular biology and biochemistry. X-ray crystallography and cryo-EM beamlines along with nuclear magnetic resonance (NMR) have informed the traditional structure-function paradigm for folded proteins for many decades. The recent advent of machine learning (ML) models for protein structure prediction, most notably AlphaFold (AF)\cite{Jumper2021,abramson2024accurate} and RoseTTAFold\cite{Baek2021}, have made stunning breakthroughs in producing accurate ground state structures of monomeric folded proteins of high quality, similar to that of experimental structures. Although static structures of folded proteins can yield significant insights, protein function is inherently controlled by dynamics.\cite{Palmer2004, Nagel2006, Fraser2009, Fenwick2014, Lane2023} But the surprising depth of deep learning models suggest that they too can create not just a single structure, but a diverse and representative protein structure-dynamical space for functional insight. 

In fact the AI field has seen an emergence of a number of approaches to learn structural variants of folded states.\cite{Lindorff2021} One expected class of ML algorithms are generative models that learn conformational distributions from molecular dynamics (MD) data via autoencoders, Boltzmann generators\cite{noe2019boltzmann} and flow-matching\cite{jing2024alphafold}, or to drive enhanced conformational sampling\cite{Degiacomi2019,Moritsugu2021} in learned lower manifolds of collective variables\cite{Vani2023}. Notably, denoising diffusion probabilistic models (DDPMs) have emerged as a powerful framework for structural variant sampling and generative protein design\cite{anand2022protein,watson2023novo,ingraham2023illuminating,hayes2024simulating,abramson2024accurate,wu2024protein,zheng2024predicting}. Compared to traditional generative approaches, DDPMs provide a more flexible mechanism to incorporate diverse conditioning signals, such as functional constraints, secondary structure preferences, or binding site requirements, enabling the generation of proteins tailored to specific tasks\cite{watson2023novo,trippe2022diffusion,ingraham2023illuminating,anand2022protein}. Other approaches involve inference time manipulations of structure prediction models. The Meiler lab cleverly demonstrated that reducing the depth of the input multiple sequence alignments (MSAs) in AF2 led to predictions of protein structures in multiple conformational states\cite{delAlamo2022}, an idea that was later elaborated upon by other groups using sequence clustering or sub-sampling the MSAs to isolate the origin of conformational states of certain classes of folded proteins.\cite{Wayment-Steele2024,Monteiro2024}  

But approximately two-thirds of the residues in the canonical human proteome are predicted to be within intrinsically disordered proteins (IDPs) or intrinsically disordered regions (IDRs) that do not adopt a dominant folded structure, but instead utilize their conformational range to execute diverse cellular functions.\cite{Wright1999,tsang2020phase} Because X-ray crystallography and cryo-EM methods are non-viable for this class of protein, computational models must bridge the gap by creating IDP ensembles that agree with the highly averaged experimental solution data from NMR, small angle X-ray scattering (SAXS), single molecule Förster Resonance Energy Transfer microscopy (smFRET) and pulsed EPR double electron-electron resonance (DEER-EPR).\cite{Bhowmick2016,Gomes2020,tesei2021deer} The generation of putative disordered ensembles can be based on atomistic molecular dynamics\cite{robustelli2018developing} or a $C_\alpha$ based coarse-grained representation\cite{cao2024coarse,tesei2024conformational}. These structural "pools" of disordered protein conformers are often reweighted to give an ensemble average that agrees with experimental observables\cite{Brookes2016, Bonomi2016, Kofinger2019, Bottaro2020, Lincoff2020}, although there is a risk that the underlying pool is incomplete and missing important populations. Recently generative algorithms have become popular for generating all-atom or coarse-grained IDP ensembles\cite{Ramanathan2021,Gupta2022,Janson2023,janson2024transferable,zhu2024precise}. For example idpGAN uses a generative adversarial network that learns a $C_\alpha$ based coarse-grained representation of disordered sequences, then superseded by a latent diffusion model with enhanced transferability to their test sequences\cite{Janson2023,janson2024transferable}. IDPFold developed by Zhu \textit{et al.} employed a two-stage training of a $C_\alpha$ backbone denoising model with PDB and all-atom molecular dynamics (MD) simulation structures, which showed promising performances on sampling structured and disordered proteins\cite{zhu2024precise}. Still they often require postprocessing by some type of reweighting step to gain better experimental relevance.  

Our group introduced a generative recurrent neural network (GRNN), DynamICE (\textbf{Dynam}ic \textbf{I}DP \textbf{C}reator with \textbf{E}xperimental Restraints)\cite{Zhang2023}, that uses experimental data to bias torsional distributions of an underlying structural pool to generate new conformational states that conform to experimental averages. The significance of the DynamICE approach is that we remove the iterative guesswork about relevant sub-populations and subsequent reweighting of arbitrarily drawn conformations. But the underlying LSTM model uses an internal coordinate representation of protein conformers, which is sub-optimal for meeting distance restraints such as nuclear overhauser effect (NOEs) and paramagnetic relaxation enhancement (PREs). Hence a method that can exploit Cartesian coordinates would be a more natural framework. Furthermore DynamICE like other Bayesian and generative models mentioned above also requires training each new IDP sequence and/or with its own experimental structural restraints, which reduces its generalizability. 

Here we shed the limitations of generic generative models and sequence specific training to create experimentally valid IDPs as well as IDRs from proteins containing folded domains, with all-atom resolution. We adapt the attention and structure modules from the ESMFold network\cite{lin2023evolutionary} to a generative model for disordered protein structures in a DDPM framework, by supplying an additional noisy structure input as exemplified in RFdiffusion\cite{watson2023novo} and AlphaFlow(ESMFlow)\cite{jing2024alphafold}. Re-purposing established networks has several benefits: it saves the considerable efforts in developing a new distributional model of accuracy and generalizability that combines sequence and inter-residue feature extraction for polymer-like biomolecules, and we enable  structure predictions of disordered regions in conjunction with folded domains with simple adjustments of diffusion steps at inference time. The resulting IDPForge (\textbf{I}ntrinsically \textbf{D}isordered \textbf{P}rotein, \textbf{FO}lded and disordered \textbf{R}egion \textbf{GE}nerator) unites ensemble generation of IDPs and IDRs with folded domains within one model, demonstrating excellent performance in sampling conformational ensembles in agreement with solution experimental data. We also introduce a sampling procedure for guidance based on experimental data to further align ensemble generation to the experimental averages that are effective for distance restraints, without additional training cost. With these diverse capabilities we expect that IDPForge can be applied across structural and integrative biology tasks for proteins that contain intrinsic disorder, and is available as an open source resource for general use.

\section{Results}
\noindent
The general framework of IDPForge is shown in Figure 1, and details of the architecture can be found in Methods. We first demonstrate the ability of IDPForge to generate single chain IDPs on 30 test sequences and assess the generated ensembles in terms of their secondary structure propensities and agreements with experimental data such as J-Couplings, nuclear Overhauser effect (NOE), paramagnetic relaxation enhancement (PRE), chemical shifts, smFRET and radius of gyration ($R_g$). We also compare IDPForge performance with other published methods, including an all-atom force field a99SB-disp\cite{robustelli2018developing} and the coarse-grained force field CALVADOS\cite{cao2024coarse}, IDPConformerGenerator\cite{teixeira2022idpconformergenerator} as representative of statistical/knowledge-based methods for generating IDPs/IDRs from sampled segments, and ML models idpGAN\cite{Janson2023} and IDPFold\cite{zhu2024precise}. In addition to the generation of single chain IDPs, we showcase two inference time generation examples with shifted distributions: 1) sampling with experimental data guidance and 2) sampling local IDRs within the context of folded domains. 

\begin{figure}[H]
    \centering
    \includegraphics[width=0.99\textwidth]{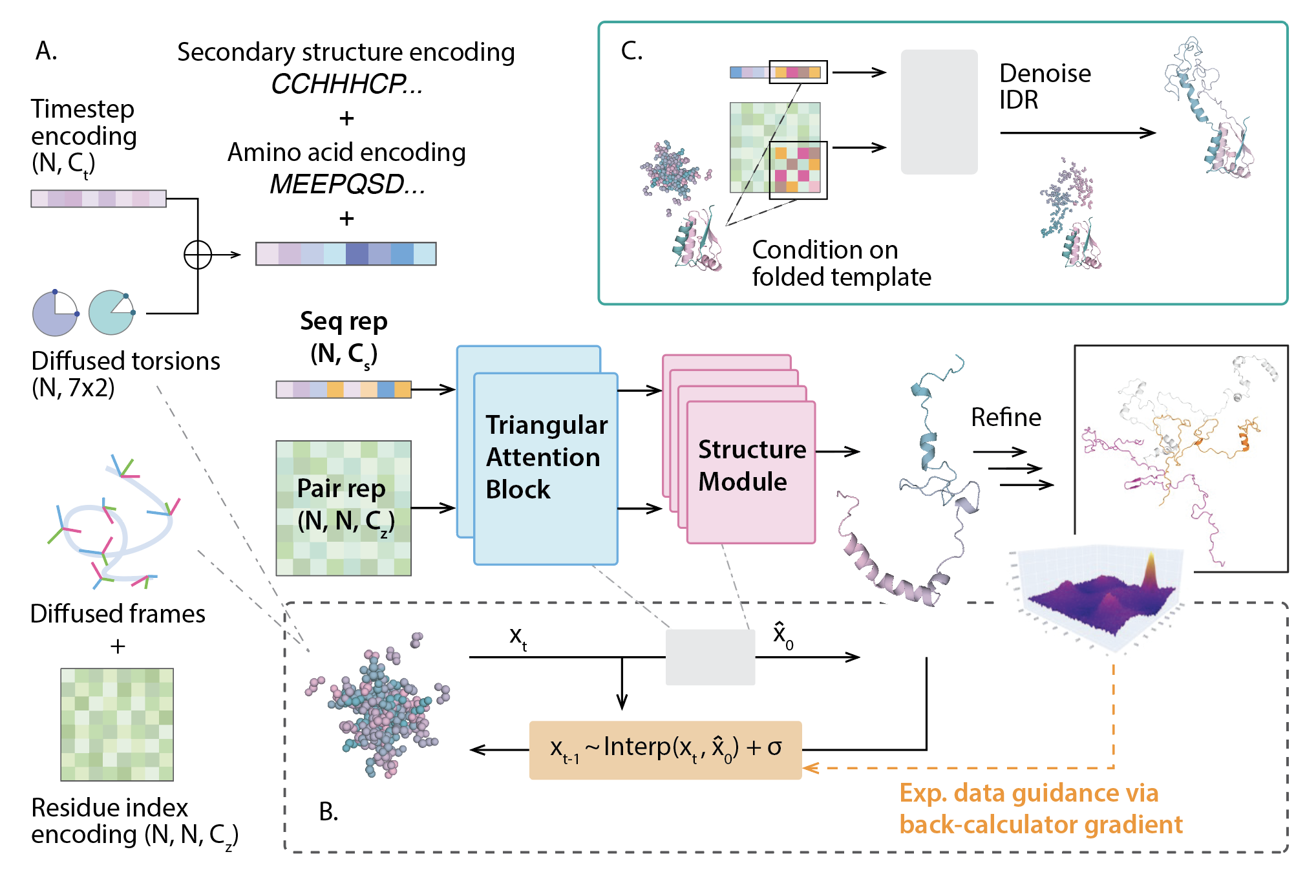}
    \caption{\textbf{The schematic of IDPForge framework for IDP/IDR prediction.} (A) Network architecture that takes a sequence and a pair representation to predict protein structure. The secondary structure encoding uses DSSP\cite{frishman1995knowledge} for helices (H) and sheets (E) assignment and further classifies DSSP coils into poly-proline II (P) and other Ramachandran regions as defined in Fig. S8. Generated structures are relaxed using an empirical forcefield. Notation: $N$ residue number, $C_t$ diffusion step encoding dimension, $C_s$ sequence representation dimension and $C_z$ pair representation dimension; $\bigoplus$ concatenation and $+$ addition. (B) Breakdown of a single denoising step $x_{t-1} \gets x_t$ into a network forward pass $\hat{x}_0 \gets \epsilon({x_t})$ and a reverse step mapping $x_{t-1} \gets \text{Interp}(x_t, \hat{x}_0)+\sigma$, with optional experimental data guidance via gradients formulated based on back-calculators. (C) Sampling of IDRs with folded domains via conditioning on a given folded template and partial denoising of the disordered sequence at inference time. Structural validation of IDPForge for bond, angle and torsion statistics over sampled conformations is provided in Supplementary Fig. 1. }
    \vspace{-5mm}
    \label{fig:network}
\end{figure}

\subsection{Generating single chain disordered ensembles}
\begin{figure}[H]
    \centering
    \includegraphics[width=0.9\textwidth]{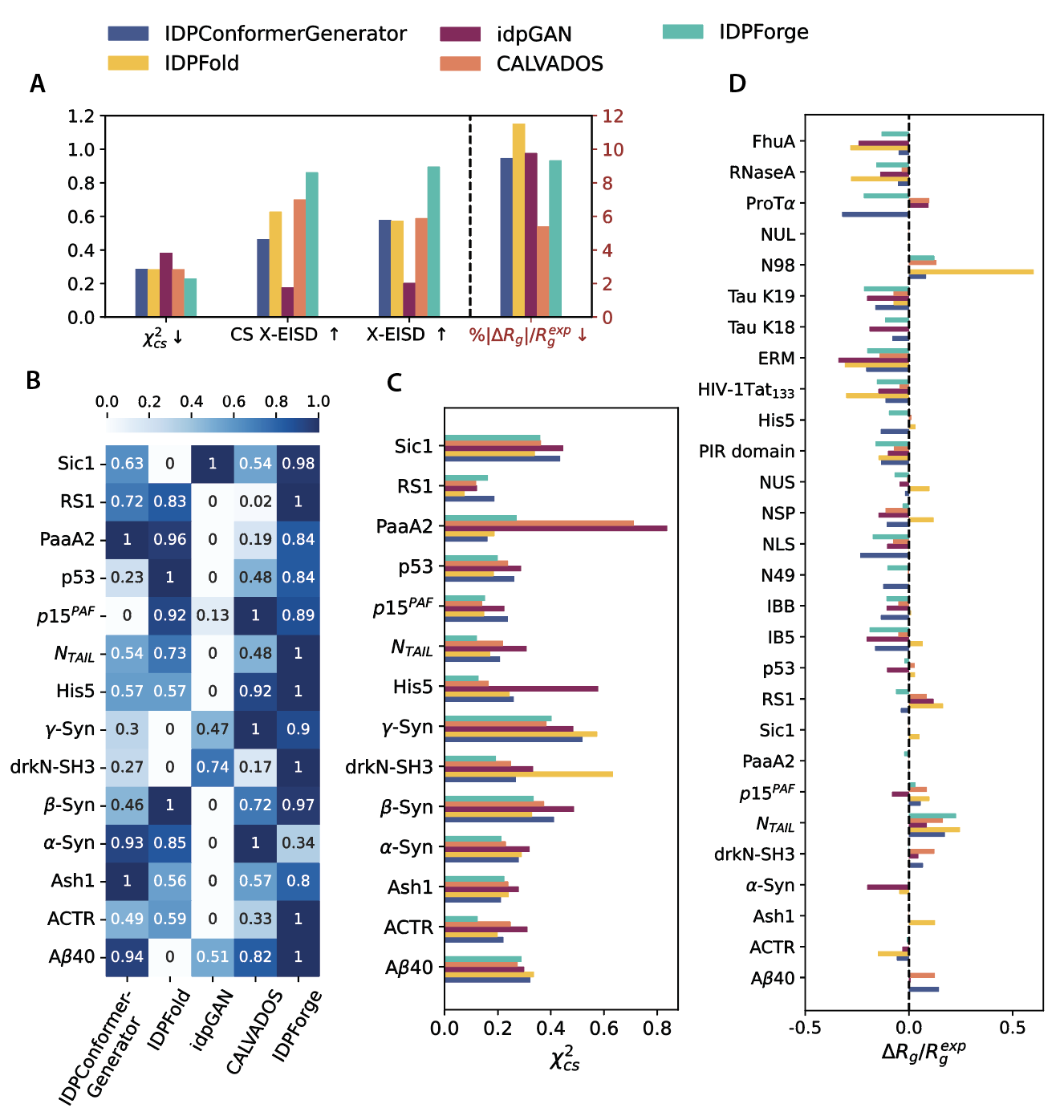}
    \caption{\textbf{Evaluation of the generated ensembles for the test sequences for various methods.} (A) Metrics used for agreement with experimental data across test IDPs. Lower $\chi^2$ and $\%|\Delta R_g|/R_g^{exp}$, and higher X-EISD scores correspond to a better experimental fit. (B) Aggregate X-EISD score over all available data types normalized per protein.\cite{Lincoff2020} (C) Chemical shift $\chi^2$ for all assigned atom types with consideration of the back-calculation errors (Eqn. S10) for 14 IDPs. (D) Normalized radius gyration error with respect to the experimental values for 28 IDPs. Experimental uncertainties were accounted for by $\Delta R_g = max(\left<R_g\right> - R_g^{exp} - \sigma^{exp}, 0) + min(\left<R_g\right>-R_g^{exp}+\sigma^{exp}, 0)$ (also see Eqn. S5). Compaction: $\Delta R_g/R_g^{exp}<0$ and expansion: $\Delta R_g/R_g^{exp}>0$. IDP sequences of test set with experimental data are provided in Supplementary Table S1. }
    \label{fig:gen_sample}
\end{figure}

\noindent
Figure \ref{fig:gen_sample} reports the test set evaluation of the generated single chain IDP ensembles across the various ensemble generation methods. These evaluations include chemical shift (CS) $\chi^2$ values for 14 sequences with CS values taken from the BMRB\cite{ulrich2007biomagresbank}, a Bayesian score using X-EISD\cite{Lincoff2020} (See Eqn. S12) for agreement with a range of smFRET experimental data and NMR data, and normalized radius of gyration error with respect to reported experiments for 28 IDPs. On average, IDPForge demonstrates excellent agreement with the solution experimental data and performs competitively in term of estimating an ensemble $\left<R_g\right>$ averaged over all test IDPs as shown in Fig. \ref{fig:gen_sample}(A). This relative performance among the methods is broken down by test evaluation category in Figs. \ref{fig:gen_sample}(B)-(D).

Fig. \ref{fig:gen_sample}(B) compares the X-EISD score for each IDP. The X-EISD score is a Bayesian model that takes into account the various sources of experimental and back-calculation uncertainties when measuring the fit between the structural ensemble and data.\cite{Brookes2016,Lincoff2020} A higher X-EISD score corresponds to a higher likelihood that the structural ensemble is in agreement with the given data. IDPForge ranks among the top by total X-EISD score across all proteins with the exception of $\alpha$-Synuclein ($\alpha$-Syn) due to a relatively high PRE error. This can be resolved by biasing the ensemble generation with PRE data and is discussed in more detail below. 

We find that the generated ensembles by all methods yield $\chi^2$ values within the uncertainties of the CS back calculator (UCBShift\cite{Li2020,Ptaszek2024}) as seen in Figure \ref{fig:gen_sample}(C), with the exception of PaaA2 and the unfolded state of drkN-SH3. Supplementary Fig. S2 sheds light on the $C_{\alpha}$ CS inconsistencies from the perspective of secondary structure features. Because PaaA2 exists in solution as two preformed helices connected by a flexible linker, the two coarse-grained methods CALVADOS and idpGAN report a higher $\chi^2$ due to their complete lack of secondary structure features. The observed connection between chemical shifts and secondary structure profile highlights that IDPs are not featureless coil-like structures. On the other hand, the drkN-SH3 ensemble generated by IDPFold features a considerable $\beta$-sheet population between residue 35-50, likely reminiscent of its folded states rather than the unfolded states\cite{zhang1995structural}, thus leading to its poor fit to the experimental chemical shifts compared to the other methods. IDPForge improves upon chemical shift predictions due to the presence of localized regions of transient secondary structure across the sequences of PaaA2 and the unfolded state of drkN-SH3. 

In regards to the $R_g$ error for the 28 test IDPs, it is found that CALVADOS performs best with IDPForge ranking second, but Fig. \ref{fig:gen_sample}(D) reveals there is a general trend for all evaluated IDP methods to systematically underestimate the $R_g$ of disordered proteins, with exceptions noted for $N_{TAIL}$ and N98 which we return to below. Finally, Supplementary Table S2 summarizes the evaluation of ensembles for 9 disordered proteins that were extracted from MD simulation trajectories using the a99SB-disp forcefield\cite{robustelli2018developing}. On this smaller test set, IDPForge shows the best aggregate X-EISD score compared to the other published methods due to smaller NOE/PRE errors, while performing as well as the a99SB-disp ensembles that report the best chemical shift and $R_g$ agreements. 

\subsection{Generation of IDP/IDR ensembles guided by experimental data}
\noindent
While IDPForge is able to generate ensembles of comparable or better performance compared to other methods for the majority of the IDPs tested in Section 2.1, we saw that all methods tend to underestimate the experimental $R_g$. Furthermore, in some cases IDP experimental information is abundant, and using it together with ensemble generation has proved fruitful for structural ensemble analysis.\cite{Gomes2020} Hence we augment IDPForge with sampling biases to further align the generated ensembles towards better agreement with the experimental data. We illustrate this on $\alpha$-Syn, whose disordered state is a precursor to toxic fibers found in Parkinson's disease\cite{Goedert2001,esteban2013identification}, and the intrinsically disordered cyclin-dependent kinase (CDK) inhibitor Sic1\cite{mittag2010structure}. 

For $\alpha$-Syn we utilize the long-range PRE experimental restraints when residue number differences are greater than 10. Table \ref{table:exp_biased} and Supplementary Fig. S3 shows that guiding the generation with PRE data alone significantly reduces the mean absolute error (MAE) by increasing the population of more extended conformers. When guiding the generation with both $R_g$ and PREs, we still observe considerable improvements in the PRE error, while the $R_g$ acts to restrain the ensembles from becoming too extended, which is corroborated by the smFRET measurement reported in Table S3. After biasing the improvements in PRE for Sic1 is driven by a heterogeneous compaction between residue 40 and the C-terminus region, and an increase in local contacts near the N-terminal as shown in Supplementary Fig. S4. 

\begin{table}[H]
\centering
\caption{\textbf{Evaluation of $\alpha-$Syn and Sic1 ensembles from unbiased and experimentally biased IDPForge compared to other methods}. We report experimental data mean absolute errors (MAEs) for paramagnetic relaxation enhancement (PRE) and ensemble $R_g$. All values are reported in terms of mean and standard deviation (in parenthesis) over 30 ensembles of 100 structures each.  PRE MAEs have factored out  experimental error according to Eqn. S5. We report MAEs for other experimental data types including J-Couplings, single molecule FRET and chemical shifts in Supplementary Table S3.}
\resizebox{0.8\textwidth}{!}{
    \begin{tabular}{lcccc}
    \toprule
     & \multicolumn{2}{c}{\textbf{$\alpha$-Syn ensembles}} & \multicolumn{2}{c}{\textbf{Sic1 ensembles}} \\
     & PRE MAE (Å) $\downarrow$ & $R_g$ (Å) & PRE MAE (Å) $\downarrow$ & $R_g$ (Å) \\
    \midrule
    IDPForge Unbiased & 7.015 & 35.53 & 2.106 & 26.83 \\
    & (0.381) & (8.37) & (0.207) & (7.27) \\ 
    IDPForge PRE biased & \textbf{2.792} & 40.93 & \textbf{1.750} & 29.77 \\
     & (0.377) & (9.17) & (0.380) & (8.12) \\ 
    IDPForge PRE+$R_g$ biased & \textbf{3.217} & 36.88 & & \\
     & (0.400) & (9.24) & & \\  
    IDPConformerGenerator & 4.028 & 34.57 & 5.133 & 27.77 \\
     & (0.291) & (7.60) & (0.615) & (5.45) \\ 
    IDPFold & 4.218 & 31.14 & 9.687 & 35.62 \\
     & (0.236) & (4.98) & (1.359) & (5.89) \\
    idpGAN & 8.511 & 26.00 & 1.820 & 26.13 \\
     & (0.388) & (6.02) & (0.230) & (6.75) \\ 
    CALVADOS & 3.689 & 37.58 & 5.816 & 28.37 \\
     & (0.147) & (7.78) & (0.574) & (4.84) \\
    a99SB-disp & 6.043 & 35.73 & 2.253 & 24.05\\
     & (0.236) & (9.51) & (0.239) & (5.83)\\
    \bottomrule
    \end{tabular}
}
\label{table:exp_biased}
\end{table}
\noindent

\vspace{-5mm}
\subsection{Sampling of IDRs along with their folded domains}
\noindent
Although most disorder in the proteome takes the form of IDRs, at present few approaches are available for modeling IDRs within the context of folded domains at all-atom resolution. Although IDPForge did not see training examples of IDRs within fixed folded domains, we test the model's ability to be conditioned on folded template structures by substituting noisy conformations with folded features and applying minimal denoising in those folded domains. We demonstrate IDPForge's zero-shot generation of terminal and linker IDRs given the structures of folded domains for two cases: modeling the missing electron density from experimentally determined IDR structures, and replacing the low confidence regions longer than 15 amino acids for IDR proteins predicted with AlphaFold2 (AF2).\cite{Jumper2021,varadi2022alphafold}

First we consider the IDRs of Interleukin-6 (IL-6) with 41 missing residues at the N-terminus\cite{shaw2014discovery}, and 
86 unresolved residues in the center of the sulfate transporter anti-sigma (STAS) domain of solute carrier family 26 member 9 (SLC26A9)\cite{chi2020structural}. Alternative to stitching based methods that screen a pre-generated pool of single chain conformations\cite{liu2023local}, the attention modules in IDPForge extract and exchange information from both the folded and noisy residue frames through pair and sequence representation for each denoising step. Hence IDPForge more intuitively enforces the spatial relation between IDRs and folded domains as well as conditioning on a folded template during generation. Figure 3 shows the generated ensembles for IL-6 and STAS from IDPForge, and Supplementary Figures S5-S7 compare the $\phi$, $\psi$ distributions, secondary structure, and RMSD of the folded domains of these same proteins when comparing CALVADOS with IDPForge. 
\begin{figure}[H]
    \centering
    \includegraphics[width=0.875\textwidth]{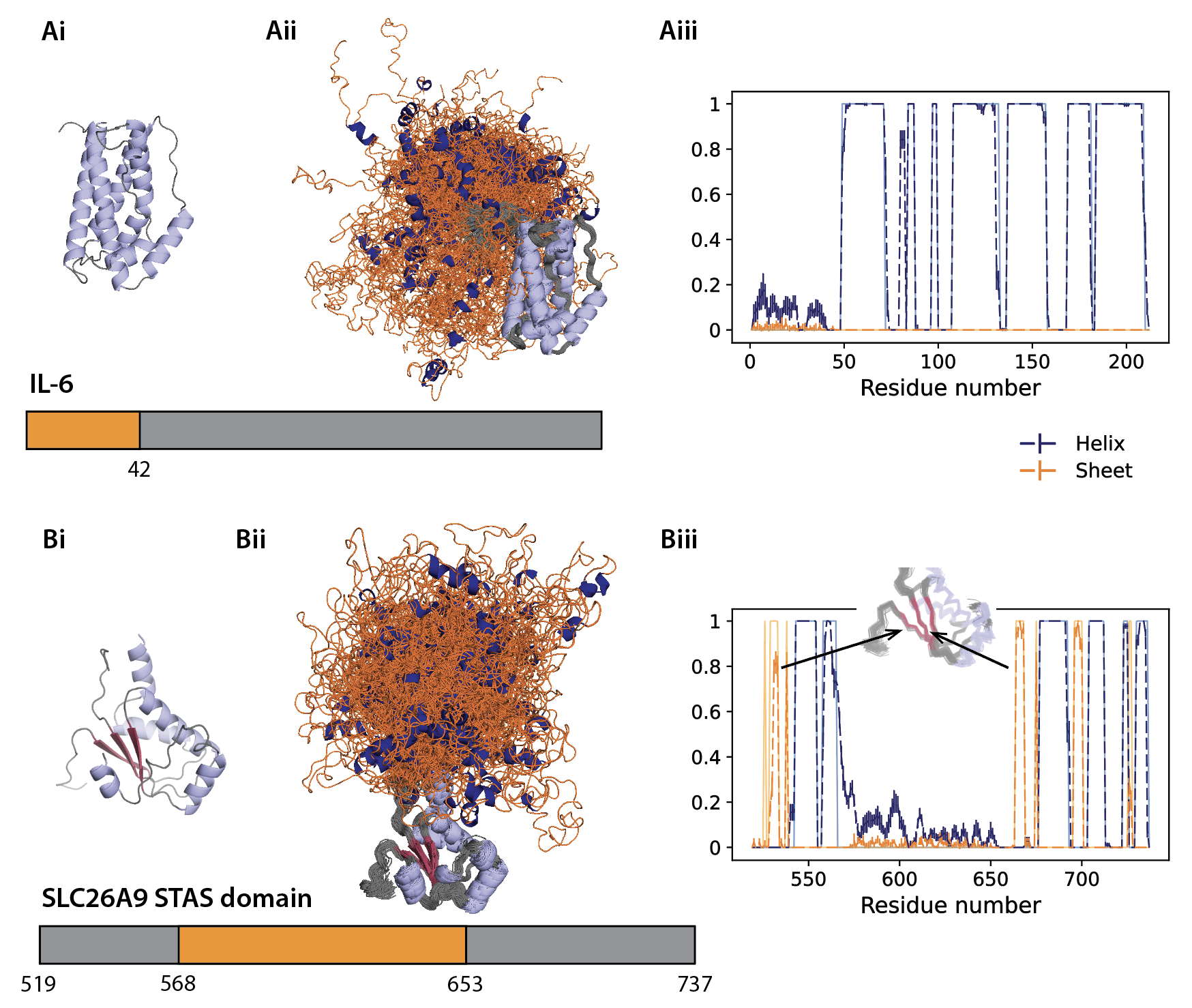}
    \caption{\textbf{All-atom IDPForge ensembles generated for IDRs with folded domains from experimental structures.} Ai) Interleukin-6 (IL-6) with 41 missing residues at the N-terminus;  PDB ID 4CNI chain C\cite{shaw2014discovery}. Bi) Intervening 85 residues of the IDR region within the sulfate transporter anti-sigma domain of solute carrier family 26 member 9 (SLC26A9 STAS domain); PDB ID 7CH1\cite{chi2020structural}. The generated ensembles of (Aii) IL-6 and (Bii) SLC26A9 with folded domains are shown in grey and IDRs in orange. Secondary structure propensities for (Aiii) IL-6 and (Biii) SLC26A9 of the generated ensembles are shown in dashed lines and the folded templates in solid and lighter colors. Errorbars estimate the mean and standard deviation of 100 conformer ensembles from 10 trials. $\beta$-sheets variations but of minor coordinate displacements are present in the close up of the modeled folded domains of SLC26A9 STAS domain. IDPForge yields RMSDs for the folded domain (core folded regions excluding 5 residues next to the disordered regions) of 0.90 Å and 1.52 Å, whereas CALVADOS has RMSDs of 1.80 Å and 2.98 Å for Il-6 and STAS, respectively.}
    \label{fig:idr_1}
\end{figure}
The structural metrics in these figures show that like CALVADOS, IDPForge preserves the secondary and tertiary structure in the folded domains of IL-6 and STAS,  whereas in the disordered regions for both IDRs it is seen that IDPForge predicts more transient secondary structure, unlike CALVADOS which is largely featureless throughout the disordered regions. In and around the sequence junction between the ordered and disordered regions, both methods exhibit more conformational variations. When the junction is helical IDPForge shows more variations across 3$_{10}$-helix, ${\alpha_R}$-helix, and $\pi$-helix states, with some transient helical structure that continues into the beginning of the disordered domain. If the junction involves a folded domain "loop", as near residue 655 in STAS, more variation is sampled. Evidence of increased conformational sampling near the sequence junctions between folded and disorder regions have been observed previously.\cite{Kim2023} 

\begin{figure}[H]
    \centering
    \includegraphics[width=0.875\textwidth]{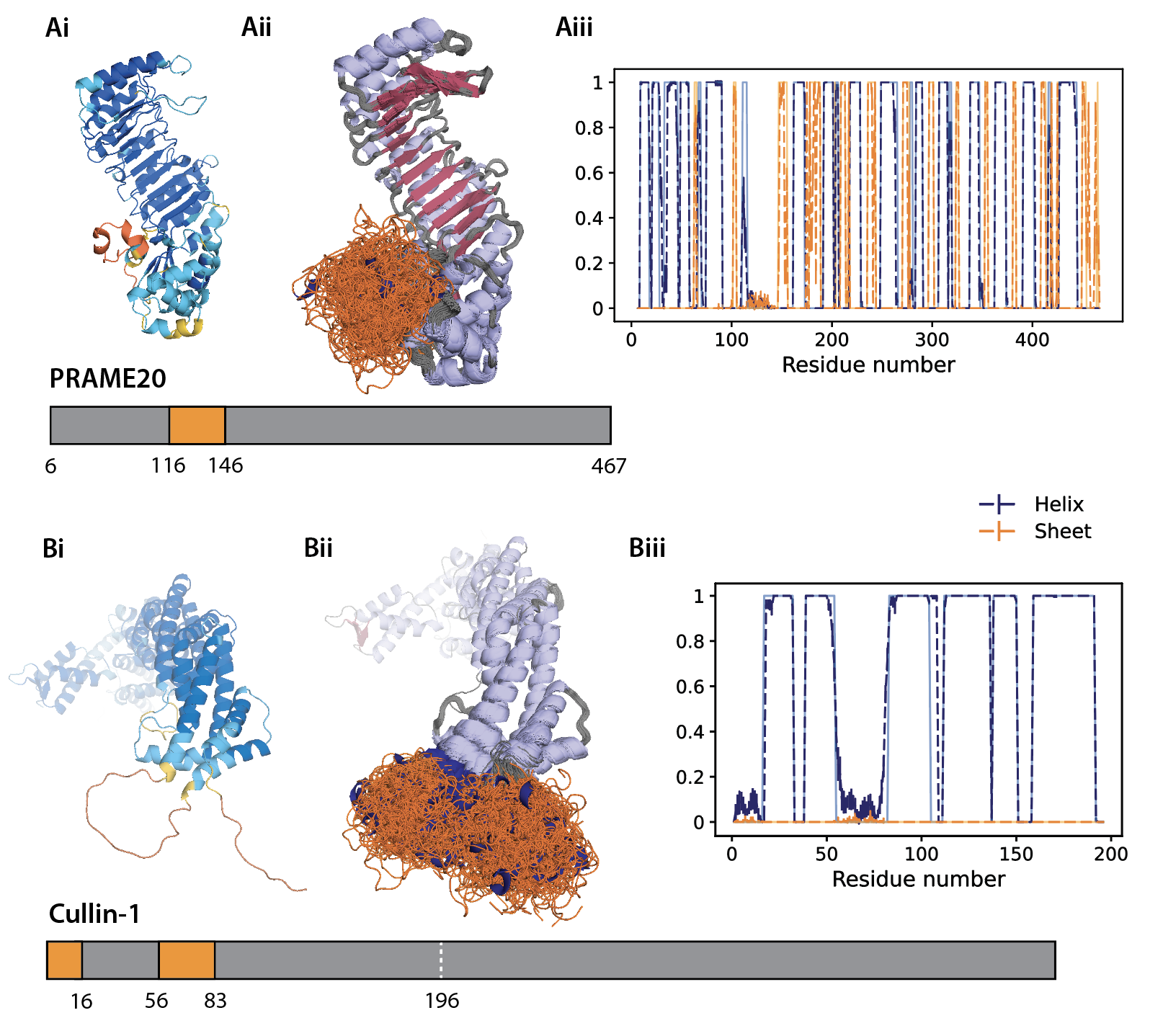}
    \caption{\textbf{All-atom IDPForge ensembles compared with AlphaFold2 predictions with IDRs representing low confidence regions.} A) PRAME family member 20 entry Q5VT98 and B) Cullin-1 entry Q13616 with AF2 predictions (i) colored by pLDDT. Yellow and orange indicate low confidence regions (pLDDT < 70). Cartoon representations (ii) of the generated ensembles with folded domains in grey and IDRs in orange. Secondary structure propensities (iii) of the generated ensembles (dashed) and the folded templates (solid and in lighter colors). IDPForge yields RMSDs for the folded domain (core folded regions excluding 5 residues next to the disordered regions) of 1.99 Å and 1.36 Å, whereas CALVADOS has RMSDs of 1.34 Å and 1.12 Å for PRAME and Cullin-1, respectively.}
    \label{fig:idr_2}
\end{figure}

In Figure 4 we consider the low confidence regions of AF2 structures of the preferentially expressed antigen in melanoma (PRAME) family member 20, replacing the low confidence region residue 116-145; we also consider Cullin-1 with two IDR regions: an N-terminal IDR 1-15 and a linker IDR between 56-82.\cite{ng1998characterization} 
We find that IDPForge conformers sample a much broader distance range than the initial AF2 models as seen in Fig. \ref{fig:idr_2} and exhibit a greater range of secondary structure types compared to the primarily featureless coil structures from AF2. Since IDPForge has transferable performance to predicting sequences containing a single IDR, we generated the 1-196 residue region for Cullin-1 by first predicting the linker IDR and holding that prediction fixed to then complete the N-terminal IDR generation. Because Cullin-1 beyond residue 196 is distant enough from the IDRs to ignore their influence on sampling, we simply aligned the generated 1-196 regions to the rest of the protein.

We also compare the IDPForge generated ensembles with the multi-domain ensembles modeled by CALVADOS in Supplementary Figs. S5-S7. We see that IDPForge conformers align closely to the backbone torsion angles from the folded templates while sampling evenly across the alpha, beta and poly-proline II in the disordered regions (Fig. S5). As already seen for the IL-6 and STAS IDRs, IDPForge creates more diverse secondary structure types for the disordered regions rather than the featureless coil structures from CALVADOS. Both IDPForge and CALVADOS sample structural variations in the folded domains with average RMSDs to template structures of $\le2$ Å, while exhibiting more flexibility at loop termini or junctions with disordered regions (Fig. S7). 

\section{Discussion}
\noindent
IDPForge is a transformer-based diffusion model that provides a conceptual advance of a united deep learning model for generating both all-atom IDPs and IDRs with folded domains. IDPForge uses attention modules for structure denoising, enabling communication between sequence and pairwise residue information while also providing a suitable channel for conditioning on folded templates. On a benchmark of 30 test IDPs/IDRs, IDPForge achieves comparable or superior agreement with NMR and smFRET experimental data compared to existing methods. It captures both local structural features and global shape characteristics as measured by the radius of gyration, and benefits from all-atom data training to more accurately represent secondary structure populations--an important factor for conforming to chemical shift data. 

Additionally, its diffusion model framework offers a natural mechanism for incorporating experimental data guidance at inference time, in which we find that all methods systematically underestimate the experimental radius of gyration. We demonstrate this capability for $\alpha$-Syn and Sic1 in which  sampling is guided with PRE data that shifts the conformational distribution towards enhanced agreement for multiple data types, including $R_g$
, with the underlying structural pool better delineating long-ranged residue interactions. We attribute the improvement to the use of the residue frame representation, which enables the model to more effectively integrate distance- and contact-biased experimental restraints—a limitation observed with the torsional representation in DynamICE\cite{Zhang2023}. We highlight that this sampling guidance strategy-requiring no training-is generalizable to any sequence with experimental data that can provide direct gradient information.

Given that a significant number of proteins contain IDRs engage in dynamic interactions with other folded domains and other IDPs/IDRs in either discrete dynamic complexes or within condensed states\cite{tsang2020phase,mittag2010protein,holehouse2024molecular}, it is valuable to extend IDPForge to model regions of disorder within the context of folded domains. We demonstrate, at inference, a targeted diffusion step strategy—initiating from a structured template instead of a fully noisy state—effectively enables this capability, even though IDPForge is not explicitly trained to perform a partial denoising task. This approach makes IDPForge particularly advantageous for interpreting linker IDRs, as it generates structures in a "geometry-aware" manner to incorporate restraints from surrounding folded regions. When conditioned on a folded template, IDPForge minimizes perturbations to the structured regions but may still introduce local changes at the junction with IDRs to maintain connectivity; some of these changes may well reflect an extension of dynamic movements from IDRs into the folded domains. Future work will explore the use of IDPForge for simultaneous modeling of multiple IDR domains and IDP/IDR complexes to further enhance its ability to capture the complex interplay between disorder and structure in biomolecular systems.


\section{Methods}
\subsection{Denoising diffusion probabilistic models.}
\noindent
Denoising diffusion probabilistic models (DDPMs)\cite{ho2020denoising} approximate a distribution by parameterizing the reversal of a discrete-time diffusion process. The forward diffusion process increasingly corrupts a sample $x_0$ from an unknown data distribution $q(x_0)$ for a sequence of $T$ steps, such that the final step $x_T$ is indistinguishable from a reference distribution that has no dependence on the data. DDPMs approximate the data distribution with a second distribution $p(x_0)$ through learning a backward transition kernel $p(x_{t-1}\mid x_t)$ at each $t$. In the reverse process, one first samples from the reference distribution $x_T \sim p(x_T)\approx q(x_T)$, and then repeatedly denoises by sampling $x_{t-1} \sim p(x_{t-1}\mid x_t)$ until $x_0 \sim p(x_0)$ is obtained. 

Under Gaussian noise $\epsilon$ and $\alpha$ that parametrizes a noise schedule (with $\bar{\alpha}_t=\prod_{s=1}^t \alpha_t$), a forward step is defined as,
\begin{equation}
    x_t=\sqrt{\bar{\alpha}_t}x_0 + \sqrt{1-\bar{\alpha}_t}\epsilon, \epsilon \sim \mathcal{N}(0, I)
\end{equation}
and the conditioned probability in the reverse process as,
\begin{equation}
    p_\theta(x_{t-1}|x_t)=\mathcal{N}(x_{t-1};\mu_\theta(x_t,t), \Sigma_\theta(x_t,t))
\end{equation}
Thus the network can learn the mean of the Gaussian posterior distribution $\mu_\theta$ from predicting the noise $\epsilon_\theta$ for the reverse process by
\begin{equation}
    \mu_\theta(x_t,t)=\frac{1}{\sqrt{\alpha_t}}(x_t-\frac{1-\alpha_t}{\sqrt{1-\bar{\alpha}_t}}\epsilon_\theta(x_t,t))
\end{equation}
which then allows us to derive a time series of $x$ by
\begin{equation}
    x_{t-1}=\frac{1}{\sqrt{\alpha_t}}(x_t-\frac{1-\alpha_t}{\sqrt{1-\bar{\alpha}_t}}\epsilon_\theta(x_t,t)) + \sqrt{1-\alpha_t}z,\quad z\sim\mathcal{N}(0, I) \text{ for } t>1
\label{eqn:reverse}
\end{equation}

\subsection{Diffusion over residue rigid frames and torsions}
\noindent 
We use DDPMs in the rigid-frame representation of residues, where the forward noising process is defined independently over the rotational and translational components of the backbone rigid bodies, as well as the torsional components of the sidechains. This treatment has been a popular choice for diffusing protein backbones in previous studies\cite{corso2022diffdock,lin2023generating,ingraham2023illuminating}. The diffusion over residue rigid frames is modeled as a discretization of a continuous-time diffusion process with Brownian motion defined on a product manifold $\mathbb{T}(3) \times SO(3)$ that separates the rotations and translations for each residue\cite{trippe2022diffusion,watson2023novo,zhu2024precise}. The translational case is simply defined by a standard Gaussian, whereas the 3D rigid body rotation $SO(3)$ can be represented by the isotropic Gaussian distribution  $\sim\mathcal{IG}_{SO(3)}$\cite{leach2022denoising, nikolayev1997normal}, which is sampled in the axis-angle parametrization with an angle of rotation $\omega$ according to
\begin{equation}
    p(\omega) =\frac{1-\cos{\omega}}{\pi}\sum_{l=0}^{\infty} (2l+1)\exp(-l(l+1)\sigma^2/2)\frac{\sin((l+1/2)\omega)}{\sin{\omega/2}}
\label{eqn:igso3}
\end{equation}
When forward noising the sidechain torsion angles with $SO(2)$, we use a wrapped normal distribution where  $wrap(\chi) = ((\chi + \pi)  mod  2\pi) - \pi$, to handle the periodic nature of torsion angles, a strategy used in FoldingDiff\cite{wu2024protein} and DiffDock\cite{corso2022diffdock}. 

While DDPM commonly models the added noise, here we trained the folding block and structure module to predict the final denoised protein structure $\hat{x}_0$ with a squared frame aligned loss\cite{jing2024alphafold}, since it was successful in sequence-to-structure prediction in AF2. We provide the pseudocode for a training step in Algorithm S1. In this case, a typical denoising step in Eqn. \ref{eqn:reverse} would include a forward network pass and a reverse step that maps the prediction $\hat{x}_0$ to $x_{t-1}$ given the noised input $x_t$ with independent backward transition kernels on the $\mathbb{T}(3) \times SO(3) \times SO(2)$ space. The translations $T_t$ and torsions $\chi_t$ are updated by interpolating the Gaussian distribution parametrized by the noise schedule $\beta_t=1-\alpha_t$; this reverse distribution for torsion angles also needs to observe the periodic boundary. On the $SO(3)$ space, the backward kernel is approximated with the score matching identity on the corresponding rotations $R_t$ and the predicted noiseless $\hat{R}_t$ such that $\nabla_{R_t}\log q_t \approx \nabla_{R_t}\log\mathcal{IG}_{SO3}(R_t;\hat{R}_0,\sigma_t^2)$. Thus to start a conformer generation, we first randomly sample $x=(T,R,\chi^{1,...,4}), ~T\sim \mathcal{N}(0, I_3), ~R\sim SO(3), ~\chi^{n=4} \sim wrap(\mathcal{N}(0, I))$ for all disordered residues, and repeatedly apply the denoising step until $T=0$ (see Algorithm S2). More details of the training scheme and sampling procedures are provided in the Supplementary Text Section 1.1. \\

\subsection{Conditioned generation towards experimental data} 
\noindent
Conditioned generative models extend the capabilities of generative algorithms by incorporating auxiliary information to meet specific data generation criteria. DDPMs provide a convenient framework for conditioned generation, as the iterative nature of the reverse process offers natural integration points to influence the trajectory of sample synthesis with auxiliary information at each step. In classifier-guided diffusion models, this guidance is achieved by directly incorporating the gradients of a classifier trained to predict the conditional label of interest $\nabla \log f_{\phi}(y|x_t)$, to steer the reverse diffusion process toward generating samples that align with condition $y$ \cite{dhariwal2021diffusion}. With Bayes' rule, we can write the log density of the joint distribution $q(x_t,y)$ as 

\begin{align}
    \nabla_{x_t}\log q(x_t,y) &= \nabla_{x_t}\log q(x_t)+\nabla_{x_t}\log q(y|x_t) \\
    &\approx-\frac{1}{\sqrt{1-\bar{\alpha}_t}}\epsilon_\theta(x_t,t)+\nabla_{x_t}\log f_{\phi}(y|x_t)
\end{align}
Introducing a guidance scale $\omega$, the classifier-guided noise prediction takes the form as,
\begin{equation}
    \bar{\epsilon}_\theta(x_t,t)=\epsilon_\theta(x_t,t)-\omega_t\sqrt{1-\bar{\alpha}_t}\nabla_{x_t}\log f_{\phi}(y|x_t)
\end{equation}

Within the context of generating protein conformers towards desired properties, this is analogous to defining a biasing potential as a function of the structural coordinates $x_{t-1}\gets x_{t-1}+\bar{\omega}_t\nabla_{x_t}P(x_t)$. Similar strategies in protein design guided by pseudo-potentials have been adopted by RFdiffusion\cite{watson2023novo} and Chroma\cite{ingraham2023illuminating} for tasks such as functional-scaffolding, generating symmetry, or shape constrained protein monomers and oligomers. Here, to bias the generation towards experimental data, we take advantage of the potentials readily defined by the back-calculators.\cite{Lincoff2020} Experimental observables such as J-Couplings, NOEs, PREs and $R_g$ (often estimated from a SAXS or smFRET measurement) have direct analytical back-calculations and we limit our studies with experimental guided generation for these data types as no training of an additional classifier is needed. We discuss details of the back-calculators in the Supplementary Text Section 1.1 and 1.3.\\

\subsection{Generation of IDRs within the context of folded domains.} 
\noindent
While IDPForge was originally trained for denoising on completely disordered single chains, we extend IDPForge to infer ensembles of local IDRs with folded domains. This involved conditioning the generation with structural featurizations of the folded domains and their secondary structure assignment in which the model effectively denoises the IDR sequences only by zero-ing the diffusion time steps of the folded domains. This partial denoising is controlled by a user defined denoising mask. Inference time modeling of IDRs has the most transferability for generating proteins with a single IDR. Thus, for the case of generating multiple IDRs separated by folded domains, we can model segments of protein containing one IDR at a time and use the denoising mask to treat previous predictions as fixed templates. Sampling of IDRs conditioned on folded domains also requires an estimation of the orientation of the folded domains with respect to the noise initiation. We prepare the folded template input by extracting its coordinates from a reference structure centered by its IDR. We randomly rotate the folded template coordinates and add a noise that scales with the residue number of the IDR to its center of mass. The detailed algorithm is provided in Algorithm S3. 

\subsection{Network architecture}
\noindent
The network architecture illustrated in Fig. \ref{fig:network} takes two main module components, the "folding block" and the structure module from ESMFold\cite{lin2023evolutionary}. The structure module consists of Invariant Point Attention (IPA) followed by backbone update network as introduced in AF2\cite{Jumper2021}. The structure module is implicitly $SE(3)$-equivariant by frame-localized geometric reasoning. The folding block contains stacks of triangular attention and communications between the sequence and pair representations that closely mimic the Evoformer described in AF2, but with the exception that it does not have a multiple sequence alignment (MSA) processing transformer module. We use two multilayer perceptrons (MLPs): one for the sequence representation to concatenate the diffused sidechain torsion angles and time step encoding, and the other for the pair representation that features a 2D transformation of the diffused coordinates before adding a pairwise relative positional encoding for the residue indices. We then define two learnable embeddings for the protein sequence and for secondary structure encoding (Supplementary Fig. S1), which are added to the sequence MLP outputs before being fed into the attention block. In an ablation study, we also investigated the effect of adding a learnable weighted sum of the ESM2 embedding to the sequence representation or a reshaped ESM2 attention map to the pair representation; we didn't find significant improvement to IDP ensemble generation performances. Details of these ESM2 appending networks are discussed in the Supplementary Text Section 1.2 and their configurations are listed in the Supplementary Table S4.

\subsection{Data preparation}
\noindent
We collected the non-overlapping disordered sequences (30-200 residues in length) from the idpGAN training set\cite{Janson2023} consisting of sequences from DisProt(version 2021-06)\cite{hatos2020disprot} and the human intrinsically disorder proteome (IDRome) database\cite{tesei2024conformational}. We excluded the native folded proteins characterized in the presence of denaturant in the original idpGAN train set that are also in the CASP12 dataset. 30 disordered sequences with experimental data (in Supplementary Table S1) and 4 sequences for systems of folded domains and IDRs (interleukin-6, SLC26A9 STAS domain, cullin-1 and PRAME20), were set aside as our test set. In addition to the disordered states, we expanded our data with CASP12 folded states from Sidechainnet\cite{king2021sidechainnet} with 30$\%$ thinning, to enrich the training data with secondary structural features. Only CASP12 sequences or segments with completely defined coordinate information and secondary structure annotations were included. We performed sequence similarity clustering with MMseqs2\cite{steinegger2017mmseqs2} and removed any sequence with $\ge 50\%$ identity to the test sequences from our train-val set, which in total generated 1,893 disordered sequences and 10112 CASP12 sequences. We used IDPConformerGenerator\cite{teixeira2022idpconformergenerator} with sequence-based sampling to generate 100 conformers for each included idpGAN sequences with FASPR\cite{huang2020faspr} sidechain building. For the IDRome sequences simulated with CALVADOS, we randomly extracted 100 conformers from the coarse-grained trajectories for each sequence, converted to all-atom conformers using cg2all\cite{pang2024simple}. All conformers were relaxed for a maximum of 5000 steps with the Amberff14SB forcefield using OpenMM\cite{eastman2023openmm}. We used MDTraj\cite{mcgibbon2015mdtraj} to assign the secondary structure for each of the generated conformers. While this yielded an extensive structural dataset of 199,540 conformers, for training efficiency a 15-3\% split was used for training-validation.

\subsection{Timing and memory usage}
\noindent
The attention modules in IDPForge are the bottleneck of scaling with $O(n^2)$ complexity, but we could use flash attention\cite{dao2022flashattention} to achieve speed ups, and also to reduce memory usage by chunking the axial attention for long sequences. IDPForge network inference using basic attention calculations on a 196 residue sequence for 40 time steps takes 2.2 seconds per conformer and relaxation takes 14.5 seconds per conformer on average on a A40 GPU. There is room for computational efficiency improvements with attention optimization and proper parallelization that we did not pursue here.

\section{Data and Code Availability}
\noindent
The code and a list of training sequences are available at https://github.com/THGLab/IDPForge.git.

\section{Acknowledgments}
\noindent
All authors acknowledge funding and thank the support from the National Institute of Health under Grant 5R01GM127627-06. J.D.F.-K. also acknowledges support from the Canada Research Chairs Program. 

\section{Author Contributions Statement}
\noindent
O.Z. and T.H.G. designed the project. O.Z. designed the deep leaning network architecture.  Z.-H. L. provided training data using IDPConformerGenerator. O.Z. and T.H.G. wrote the paper. All authors discussed the results and made comments and edits to the manuscript.

\section{Competing Interests Statement}
\noindent
The authors declare no competing interests.

\bibliography{references}
\bibliographystyle{naturemag}


\end{document}